\documentclass[preprint,aps,amsmath,nofootinbib]{revtex4}
\pdfoutput=1
\usepackage{graphicx}
\usepackage{bm}
\usepackage{graphics}
\usepackage{amsfonts,amssymb}
\usepackage{color}
\usepackage{hyperref}

\newif\ifpdf
\ifx\pdfoutput\undefined
\pdffalse 
\else
\pdfoutput=1 
\pdftrue
\fi

\def\Dsl{\hbox{/\kern-.6000em D}} 

\def\dsl{\,\raise.15ex\hbox{/}\mkern-13.5mu D}

\def\ltap{\ \raise.3ex\hbox{$<$\kern-.75em\lower1ex\hbox{$\sim$}}\ }
\def\gtap{\ \raise.3ex\hbox{$>$\kern-.75em\lower1ex\hbox{$\sim$}}\ }
\def\OMIT#1{}

\def\lsim{\mathrel{\raise.3ex\hbox{$<$\kern-.75em\lower1ex\hbox{$\sim$}}}}
\def\gsim{\mathrel{\raise.3ex\hbox{$>$\kern-.75em\lower1ex\hbox{$\sim$}}}}

\def\msb{{\overline{\rm MS}}}

\newcommand{\nn}{\nonumber}

\newcommand{\ord}{{\cal O}}

\def\msb{{\overline{\rm MS}}}

\catcode`\@=11
\def\slash{\mathpalette\make@slash}
\def\make@slash#1#2{\setbox\z@\hbox{$#1#2$}%
  \hbox to 0pt{\hss$#1/$\hss\kern-\wd0}\box0}
\catcode`\@=12 

\newcommand{\be}{\begin{equation}}
\newcommand{\ee}{\end{equation}}
\newcommand{\bea}{\begin{eqnarray}}
\newcommand{\eea}{\end{eqnarray}}

\definecolor{orange}{rgb}{1,0.5,0}

\newcommand{\sss}{\scriptscriptstyle}

\begin{document}


\preprint{ \vbox{ 
\hbox{UWThPh-2012-27}
\hbox{DESY 12-156}  
}}

\title{\phantom{x}\vspace{0.5cm} 
NLL resummation for the static potential in ${\cal N}=4$ SYM theory\\
\vspace{1.0cm} }

\author{
 Maximilian~Stahlhofen\footnote{Electronic address: maximilian.stahlhofen@desy.de}}
\affiliation{
University of Vienna, Faculty of Physics, Boltzmanngasse 5, A-1090 Wien, Austria \vspace{0.5cm}\\
DESY Theory Group, Notkestra\ss e 85, D-22607 Hamburg, Germany
\vspace{1.5 cm}
}


\begin{abstract}
\vspace{0.5cm}
\setlength\baselineskip{18pt}
We determine the complete NLL running of the static potential associated with the locally 1/2 BPS Wilson loop in ${\cal N}=4$ supersymmetric Yang-Mills theory. We present results for the $SU(N_c)$ singlet as well as for the adjoint configuration and arbitrary $N_c$ at weak coupling. In order to derive the respective anomalous dimensions we perform a two-loop calculation in the ${\cal N}=4$ supersymmetric version of the effective field theory pNRQCD. In addition we confirm the recently obtained fixed-order result for the singlet static potential generated exclusively by ladder diagrams to the third order in the t`Hooft coupling.
We also give an explicit expression for the logarithmic contribution of all non-ladder diagrams at this order.
\end{abstract}
\maketitle


\newpage

\section{Introduction}
\label{sectionintro}

${\cal N}=4$ supersymmetric Yang-Mills (SYM) theory in four spacetime dimensions is commonly believed to be integrable in the large-$N_c$ (planar) limit.
The reason is connected to the conjectured duality between certain pairs of conformal quantum field and string theory models, the so called AdS/CFT correspondence~\cite{Maldacena:1997re,Gubser:1998bc,Witten:1998qj}. Notably it maps observables in planar ${\cal N}=4$ SYM theory to dual observables in type IIB superstring theory on an $AdS_5 \times S^5$ background with vanishing string coupling. This string theory in turn belongs to a class of two dimensional (worldsheet) models, which appear to be integrable~\cite{AdSCFTIntegrability}.

The key feature of ${\cal N}=4$ SYM theory is its large symmetry content and in particular its conformal symmetry, which is necessary for the AdS/CFT duality to apply. As a consequence of the conformal invariance the beta function of the ${\cal N}=4$ SYM gauge coupling $g$ vanishes to all orders. 
Correspondingly also the t`Hooft coupling $\lambda \equiv N_c\, g^2$, which is kept finite in the large-$N_c$ model, is scale independent.

Despite the assumed integrability finding exact solutions in terms of algebraic functions of the coupling constant for observables like scattering amplitudes or expectation values of gauge invariant Wilson loops in planar ${\cal N}=4$ SYM theory is far from trivial.
A strong motivation to continue the effort in this direction is the hope that some (qualitative) features of such solutions are universal for a wider class of four-dimensional gauge theories including more realistic ones like QCD. Evidence for some universal behaviour has e.g. been found for gluon scattering amplitudes. Likewise in some instances ${\cal N}=4$ SYM results might possibly be regarded as the first approximation of the respective results in less symmetric Yang-Mills theories. On the other hand it is also conceivable to get new insigths into string theory/quantum gravity by studying exact results derived  by means of field theory methods in ${\cal N}=4$ SYM theory due to the AdS/CFT correspondence.
In order to formulate and test an all-orders ansatz for an observable in (planar) ${\cal N}=4$ SYM theory it is crucial to know the perturbative results at weak and strong coupling. 
The first can be computed within perturbative quantum field theory, the second within perturbative string theory exploiting the AdS/CFT duality.

In this work we consider the energy of two static sources, which are in the fundamental (quark) and antifundamental (antiquark) representation of $SU(N_c)$, respectively, and are separated by the spatial distance $r$. The system can either be in an overall $SU(N_c)$ singlet or an overall $SU(N_c)$ adjoint multiplet configuration, the latter of which interacting with (external) ${\cal N}=4$ SYM fields to form a physical quantity.
In particular we focus on the singlet static energy given by
\begin{align} 
E_s(r)=\lim_{T \rightarrow \infty}\frac{i}{T}\ln \langle W_\Box \rangle\,,
\label{Es}
\end{align}
where $W_\Box$ is a rectangular representative of the class of 1/2 BPS Wilson loops
\begin{align}
W_C=\frac{1}{N_c}\,{\rm Tr}\; {\cal P}\, \exp \Big( -ig\oint_C d\tau(A_{\mu}{\dot x}^{\mu}+\Phi_ n|{\dot x}|) \Big)
\label{WC}
\end{align}
with the path $C=\Box$ spanned by the spacetime points $x_1 = (T/2,{\bf r}/2)$, $x_2 = (T/2,-{\bf r}/2)$, $y_1 = (-T/2,{\bf r}/2)$ and  $y_2 = (-T/2,-{\bf r}/2)$.
The definition of the Wilson loop, Eq.~\eqref{WC}, is motivated in Ref.~\cite{Drukker:1999zq} with the static quarks modeled by W-bosons. We will often refer to the corresponding energy in Eq.~\eqref{Es} as the static potential. It is among the most prominent observables studied in the context of gauge/string duality.
The main difference to the QCD analog is that the sources not only couple to gluons, but also to the six scalar fields $\Phi_i$ of four-dimensional ${\cal N}=4$ SYM theory through the $\Phi_n \equiv \bf \Phi \cdot {\hat n}$ term in Eq.~\eqref{WC}, ${\bf \hat n}$ being a six-dimensional unit vector (${\bf \hat n}^2 =1$).

The perturbative calculation of the ${\cal N}=4$ SYM static potential in terms of quark-antiquark interaction diagrams relies on the (full theory) Lagrangian\footnote{The analogy of the static potential to a quark-antiquark scattering process on the diagram level in momentum space is actually only evident for a number of gauges including the generalized Lorenz gauge, cf. Ref.~\cite{SchroederPhD} for the QCD case. We will therefore restrict ourselves to the latter in the following discussion.}
\begin{align}
 {\cal L}={\cal L}_{stat}+{\cal L}_{{\cal N}=4}\,.
\label{Lagrangian}
\end{align}
Eq.~\eqref{WC} fixes the interaction terms of the static (quark) sources $\psi$ and $\bar \chi$ in the fundamental and antifundamental representation, respectively,
\begin{align}
{\cal L}_{stat}= \psi^{\dagger}(i\partial_0-gA_0-g\Phi_n)\psi+
{\bar \chi}^{\dagger}(i\partial_0+gA_0^{T}-g\Phi^{T}_n){\bar \chi}\,,
\label{Lstat}
\end{align}
and ${\cal L}_{{\cal N}=4}$ denotes the usual Lagrangian of ${\cal N}=4$ SYM theory without external sources in four-dimensional Minkowski space~\cite{Gliozzi:1976qd,Brink:1976bc},
\begin{align}
{\cal L}_{{\cal N}=4} =& - \frac14 F_{\mu \nu}^{a} F^{\mu \nu , a}
+
\frac12 \sum_{i=1}^6 (D_{\mu} \Phi_i)^a (D^{\mu}\Phi_i)^a + \frac12 {\bar \Psi_j}^a \gamma_{\mu} (D^{\mu}\Psi_j)^a 
+ \cdots\,.
\label{LN4}
\end{align}
The $\Psi_j$ ($j=1,..,4$) are four Majorana four-spinors living in the adjoint representation of $SU(N_c)$ just like the gluon field $A^\mu$ and the scalar field $\Phi_i$ ($i=1,..,6$). For the gauge covariant derivative acting on these fields we adopt the following (sign) convention: $(D_\mu X)^a = \partial_\mu X^a - g f^{abc}A^b_\mu X^c$, where $X=A^\mu,\Phi_i,\Psi_j$.

A number of perturbative calculations of the static potential in (planar) ${\cal N}=4$ SYM theory in the weak as well as in the strong coupling regime can be found in the literature.
For $\lambda \gg 1$ present string theory calculations~\cite{Chu:2009qt,Forini:2010ek} provide the first perturbative quantum correction of relative $\ord(\lambda^{-1/2})$ to the classical string result~\cite{Maldacena:1998im,Rey:1998ik}. See also Ref.~\cite{Drukker:2011za} for smooth generalizations of the rectangular Wilson loop to other shapes like the circular Wilson loop.

A recent calculation of the ${\cal N}=4$ SYM static potential at weak coupling ($\lambda \ll 1$) considers diagrams of ladder type only~\cite{Correa:2012nk}. Besides the explicit result up to (absolute) $\ord(\lambda^3)$ Ref.~\cite{Correa:2012nk} also shows how a summation of all ladder diagrams can be achieved by solving a simple Schr\"odinger problem. 
In Ref.~\cite{Henn:2012qz} an algorithm is presented, which returns the perturbative solution to the Schr\"odinger problem for the more general case of the cusp anomalous dimension at any finite order in the loop expansion. The ladder diagrams for the static potential with an arbitrary number of loops, i.e. with an arbitrary number of ``ladder rungs'', can be obtained as a special case.\footnote{The cusp anomalous dimension is equivalent to the static quark-antiquark potential on the sphere $S^3$, where the quarks are separated by an angle $\delta$~\cite{Correa:2012nk}. The limit $\delta \to 0$ gives the static potential in flat space.}

While ladder diagrams are all there is at the tree, $\ord(\lambda)$, and one-loop level, $\ord(\lambda^2)$~\cite{Erickson:1999qv,Erickson:2000af}, diagrams with other topologies contribute at two loops, $\ord(\lambda^3)$, and beyond\footnote{Note that in contrast to Ref.~\cite{Correa:2012nk} we stick to the standard counting of loops in QCD (scattering) diagrams.}. In Ref.~\cite{Bykov:2012sc} a systematic generalization of the results of Ref.~\cite{Correa:2012nk} in terms of an expansion around the ladder result for the static potential is discussed. As a first step the authors computed the leading logarithmic correction to the ladder limit for weak and strong coupling. Ref.~\cite{Henn:2012qz} further developes and generalizes this approximation method to the cusp anomalous dimension.

In the present work we are concerned with the weak coupling limit of the ${\cal N}=4$ SYM static potential. 
The strictly perturbative approach in terms of on-shell amplitudes derived from the Lagrangian in Eq.~\eqref{Lagrangian} is spoiled by infrared (IR) singularities. They occur first at one loop and do not cancel among each other in the individual terms of the loop expansion. A similar problem in the perturbation series of the QCD static potential has been known for a long time. In four spacetime dimensions the perturbative QCD corrections start to exhibit IR divergences at the three-loop~\cite{Appelquist:1977es} and in three dimensions at the two-loop level~\cite{SchroederPhD}. The reason for the divergent behaviour in the IR is that the strictly perturbative approach fails to allow for effects at the so-called ``ultrasoft'' scale $\sim \lambda/r$. This scale parametrizes the difference between the binding energy of (intermediate) adjoint and singlet states of the quark-antiquark system.
The ultrasoft scale does not appear in single quark-antiquark interaction diagrams. It is rather generated by a resummation (exponentiation) of certain contributions of loop diagrams to all orders and acts as a physical IR regulator.

In Ref.~\cite{Correa:2012nk} a corresponding resummation is explicitly performed for ladder diagrams and finite results through next-to-next-to-leading order (NNLO), i.e. $\ord(\lambda^3)$, are obtained. These results are now non-analytic in the coupling constant $\lambda$ since they contain logarithms of the ratio between the soft scale $\sim 1/r$ and the ultrasoft scale $\sim\lambda/r$.\footnote{The terminology for the scales (soft, ultrasoft) actually comes from the more physical situation of dynamical quarks with a mass, which is large but finite and defines a third so called ``hard'' scale $m\gg1/r\gg \lambda/r$.} They lead to the typical $\ln \lambda$ terms, which are strongly connected to the logarithmic IR singularities of the strictly perturbative calculation.

A more systematic solution to the problem of IR singularities in the QCD static potential has been proposed in Ref.~\cite{Brambilla:1999qa} and has been successfully applied not only in four, but also in three spacetime dimensions~\cite{Pineda:2010mb}. It is based on the effective field theory (EFT) potential nonrelativistic QCD (pNRQCD)~\cite{pNRQCDfirst} (for a review see Ref.~\cite{pNRQCDreview}). This framework explicitly contains degrees of freedom living at the ultrasoft scale, most prominently ultrasoft gluons. It allows to systematically calculate the ultrasoft contributions missing in the strictly perturbative calculation from a finite number of EFT diagrams at every order in the coupling constant. Most notably the ultraviolet divergences of the pNRQCD results exactly cancel the IR divergences from the respective full theory (QCD) diagrams, which (in the static limit) only involve the soft scale $1/r$.

The analogous EFT for ${\cal N}=4$ SYM theory was introduced by Pineda in Ref.~\cite{Pineda:2007kz}. Also the complete next-to-leading order (NLO), i.e. $\ord(\lambda^2)$, result of the singlet static energy $E_s(r)$ at weak coupling was derived there for the first time. The result also holds for arbitrary $N_c$. In addition, Ref.~\cite{Pineda:2007kz} makes use of the renormalization group (RG) formalism to determine the leading logarithmic (LL) running of the static potential from the leading ultrasoft UV divergence in the planar limit. To some extent this already represents an all-orders result and demonstrates the power of the EFT approach.

It is the aim of the present work to compute the next-to-leading logarithmic (NLL) RG evolution of the static potential in this EFT framework for arbitrary $N_c$. With the result we will be able to deduce also the full NNLO expression for the singlet static potential in the planar limit, including the contributions from non-ladder diagrams, up to a (soft) constant and thus generalize the results of Refs.~\cite{Correa:2012nk,Bykov:2012sc}.

The outline of this paper is as follows.
In Sec.~\ref{sectionEFTCalc} we briefly review the theoretical framework we are working in and explain the ultrasoft two-loop calculation we have performed. Explicit results for individual diagrams are given in App.~\ref{twoloopdiags}. In Sec.~\ref{sectionRG} we solve the renormalization group equations (RGE's) derived in Sec.~\ref{sectionEFTCalc} allowing for the NLO soft matching conditions for the singlet and adjoint static potentials. We then give the complete NLL results for the respective static energies. Section~\ref{sectionlargeNc} is concerned with the large-$N_c$ limit of our results. After computing also the soft two-loop matching condition from ladder graphs we finally present the full large-$N_c$ NNLO result for the singlet static energy up to an unknown constant coming from soft two-loop non-ladder diagrams. We conclude in Sec.~\ref{sectionConclusion}.

\section{The EFT Calculation}
\label{sectionEFTCalc}

The EFT we are employing in this work was introduced in Ref.~\cite{Pineda:2007kz} and is based on pNRQCD. It exploits the hierarchy between the soft and the ultrasoft scale in the weak coupling limit, $E_s\sim \lambda/r \ll 1/r$ and makes the factorization between these scales manifest. 
This is achieved by the ``multipole'' expansion of the effective Lagrangian in powers of the distance $r$.
Fluctuations with energies and/or momenta $\sim 1/r$ are integrated out. The remaining degrees of freedom are massless gluons, scalars and fermions with ultrasoft energies and momenta and the quark-antiquark pair with ultrasoft energy and zero momentum. The latter is conveniently projected on the physical singlet (S) and adjoint (O) states. In analogy to QCD we will often refer to the $SU(N_c)$ adjoint as the ``octet'' in the following.

At leading order in the multipole expansion the singlet-octet sector of the EFT Lagrangian~\cite{Pineda:2007kz} reads
\begin{align}
{\cal L}_{\rm US} =& \; {\rm Tr} \bigg\{ {\rm S}^\dagger \left( i\partial_0  - V_s(r) \right) {\rm S}  + {\rm O}^\dagger \left( iD_0 - V_o(r) \right) {\rm O} \bigg\}
\nn\\
& -2 g V_{\Phi}(r){\rm Tr}\left\{{\rm S}^{\dagger}\Phi_n {\rm O}+{\rm S}\Phi_n {\rm O}^{\dagger}\right\}  
  - g V_{\Phi_O}(r){\rm Tr}\left\{{\rm O}^{\dagger}\left\{\Phi_n, {\rm O}\right\}\right\}
  + \ord(r)\,.
\label{LUS}
\end{align}
S and O are operators in color space and related to the normalized singlet and octet fields by
\begin{align}
{\rm S} \equiv \frac{ 1\!\!{\rm l}_c }{ \sqrt{N_c}} S\,, \qquad {\rm O} \equiv  \frac{ T^a }{ \sqrt{T_F}}O^a, 
\label{norm}
\end{align}
where the $T^a$ are the generators of the fundamental representation of $SU(N_c)$ ($T_F=1/2$). The EFT Lagrangian of course respects all symmetries of the full theory Lagrangian, Eq.~\eqref{Lagrangian}.
The terms in Eq.~\eqref{LN4} retain their form for the ultrasoft gluons, scalars and fermions. Accordingly the Feynman rules for the interaction of these fields among each other are the same as in full ${\cal N}=4$ SYM theory. The EFT Feynman rules from the Lagrangian in Eq.~\eqref{LUS} are listed in Appendix~\ref{feynrules}.

The Wilson coefficients $V_i$ depend on the distance $r$ and on the renormalization scale $\nu$. They are typically fixed at the soft scale ($\nu \sim 1/r$) by matching to the full theory. We will refer to the coefficients
\begin{align}
V_s(r) \equiv  - 2C_F \frac{\alpha_{V_s}(r)}{r}, 
\qquad
V_o(r) \equiv  2\left(\frac{C_A}{2} -C_F\right) \frac{\alpha_{V_o}(r)}{r},
\label{defpot}
\end{align}
as the (soft) singlet and octet static potentials, respectively ($C_A=N_c$, $C_F=(N_c^2-1)/(2N_c)$). 
Tree level matching gives $\alpha_{V_s} = \alpha_{V_o} = \alpha \equiv \frac{g^2}{4\pi}$, $V_{\Phi}=V_{\Phi_O}=1$.
The potential $V_{s/o}$ is to be distinguished from the observable static energy\footnote{As mentioned above we will sometimes also use the term ``potential'' for the static energy. We will however be more precise, whenever the distinction between the Wilson coefficient $V_{s/o}$ and the observable energy $E_{s/o}$ is important.}
\begin{align}
 E_{s/o}(r) = V_{s/o} + \delta E_{s/o}^{us}\,.
\label{EsEFT}
\end{align}
The Wilson loop definition of the static octet energy differs from the singlet energy, Eq.~\eqref{Es}, by an insertion of the $SU(N_c)$ generator $T^a$ on each of the spatial lines of the rectangular loop, see e.g. Ref.~\cite{pNRQCDreview} for the QCD analog.
$\delta E^{us}$ is the ultrasoft contribution to the static energy. We will compute it below for the singlet as well as for the octet at the two-loop level.
Eq.~\eqref{EsEFT} holds for renormalized as well as for bare quantities on the right hand side.

In calculations of $\delta E^{us}$ the ultrasoft scale is represented by $\Delta V \equiv V_o - V_s$. In fact the potential coefficients $V_s$ and $V_o$ only appear in this combination.
At NLO in the multipole expansion the operators of ${\cal L}_{\rm US}$ are proportional to the the distance vector ${\bf r}$ and come with at least one power of the coupling constant $g$, see Ref.~\cite{Pineda:2007kz}. Since rotational invariance requires an even number of NLO operators their contribution to the static energy scales like $\alpha \Delta V^3 {\bf r^2} \sim \alpha^4/r$. The NNLO operators ($\propto \alpha {\bf r^2}$) start to contribute at the same parametric order. This is beyond the precision we are aiming for in the present work. We can therefore safely neglect operators of dimension greater than four and work only with the LO operators in Eq.~\eqref{LUS}.

Let $-i \Sigma_{s/o}(E)$ denote the sum of all singlet/octet 1PI self-energy diagrams in the EFT. Then the full propagator of the interacting singlet/octet field takes the form
\begin{align}
 \frac{i}{E-V_{s/o}^{bare} - \Sigma_{s/o}(E)} \;\xrightarrow{E \,\to\, E_{s/o}(r)}\; \frac{i Z_{s/o}}{E-E_{s/o}(r)}\,.
\label{propagator}
\end{align}
The arrow indicates the physical on-shell limit and $\sqrt{Z_{s/o}}$ is the singlet/octet wave function renormalization constant.
In the following quantities without the ``bare'' label are understood to be renormalized.
From Eqs.~\eqref{EsEFT} and~\eqref{propagator} one finds 
\begin{align}
\delta E_{s/o}^{us} &= Z_{s/o}\,\big(\delta V_{s/o} + \Sigma_{s/o}(E\!=\!V_{s/o})\big) \,+\, \ord(\alpha^3)\,, \label{deltaEus} \\[2 ex]
Z_{s/o} &= \big(1 - \Sigma_{s/o}'(E\!=\!V_{s,o})\big)^{-1} \,+\, \ord(\alpha^2)
\end{align}
to second order in the loop expansion. The term $\Sigma_{s/o}'(E\!=\!V_s)$ denotes the derivative of the self-energy $\Sigma_{s/o}$ with respect to the external energy $E$ evaluated at $E = V_{s/o}$ and $\delta V_{s/o}=V_{s/o}^{bare}-V_{s/o}$ is the counterterm of the singlet/octet potential in a given renormalization scheme. 

Before we turn to the actual calculation of $\delta E_{s/o}^{us}$ we first consider possible higher order corrections to the Wilson coefficients $V_\phi$ and $V_{\phi_O}$ as they can mix into the result for the static energy. The coupling constant $g$ does not receive corrections due to conformal invariance inherited from the full theory.\footnote{We have checked this statement explicitly at one loop level in the EFT.} It has been stated already in Ref.~\cite{Pineda:2007kz} that there are no corrections to $V_\phi$ and $V_{\phi_O}$ at $\ord(\alpha)$. Indeed we have found vanishing LL anomalous dimensions by direct computation in the EFT. Inspecting the soft one-loop diagrams relevant for the $\ord(\alpha^2)$ matching of $V_\phi$ and $V_{\phi_O}$ one finds moreover that they possibly contain IR linear but no logarithmic divergences.
As a consequence also the NLL anomalous dimensions must vanish. Summarizing we have ($n=0,1,2,\dots$)
\begin{align}
V_\phi = 1 + \ord\big(\alpha^2 (\alpha \ln \alpha)^n \big)\,,\qquad
V_{\phi_O} = 1 + \ord\big(\alpha^2 (\alpha \ln \alpha)^n \big)\,.
\end{align}
Thus mixing corrections to Eq.~\eqref{deltaEus} are higher order and we will consistently use $V_\phi = V_{\phi_O}= 1$ in the following.

Throughout this paper we will work in Feynman gauge using dimensional regularization ($d=4 -2\epsilon$) and the $\msb$ renormalization scheme, unless explicitly stated otherwise.
We parametrize our results for the ultrasoft contributions to the static energies and the anomalous dimensions of the potentials as follows
\begin{align}
 \delta E_{s}^{us} &= - C_F \Delta V \, \Big(\, \frac{\alpha}{\pi}\, b_s^{(1)} \,+\, \frac{\alpha^2}{\pi^2}\, b_s^{(2)} \,+\, \ord(\alpha^3) \,\Big) \,+\, \ord(r^2)\,, 
\label{EusSing} \\
 \delta E_{o}^{us} &= \Big(\frac{C_A}{2} - C_F\Big) \Delta V \, \Big(\, \frac{\alpha}{\pi}\, b_o^{(1)} \,+\, \frac{\alpha^2}{\pi^2}\, b_o^{(2)} \,+\, \ord(\alpha^3) \,\Big) \,+\, \ord(r^2)\,, 
\label{EusOct} \\[0.4 cm]
\nu \frac{d}{d\nu}\, V_{s} &= - C_F \Delta V \, \Big(\, \frac{\alpha}{\pi}\, c_s^{(1)} \,+\, \frac{\alpha^2}{\pi^2}\, c_s^{(2)} \,+\, \ord(\alpha^3) \,\Big) \,+\, \ord(r^2)\,, 
\label{SingAnomDim} \\
\nu \frac{d}{d\nu}\, V_{o} &= \Big(\frac{C_A}{2} - C_F\Big) \Delta V \, \Big(\, \frac{\alpha}{\pi}\, c_o^{(1)} \,+\, \frac{\alpha^2}{\pi^2}\, c_o^{(2)} \,+\, \ord(\alpha^3) \,\Big) \,+\, \ord(r^2)\,.
\label{OctAnomDim}
\end{align}
The latter are determined from the potential counterterms in Eq.~\eqref{deltaEus} in the standard way,
\begin{align}
 \nu \frac{d}{d\nu}\, V_{s/o} = - \nu \frac{d}{d\nu} \,\delta V_{s/o}\,,
\end{align}
using $\nu \,d\alpha/d\nu = - 2 \epsilon \alpha$.
Note that for large $N_c$ the octet expressions in Eqs.~\eqref{defpot}, ~\eqref{EusOct} and~\eqref{OctAnomDim} are suppressed by two powers in the $1/N_c$ expansion as compared to the singlet. 

\begin{figure}[t]
\includegraphics[width=0.25 \textwidth]{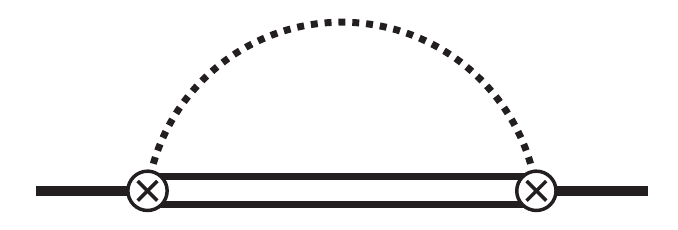}
\qquad\qquad\qquad
\includegraphics[width=0.25 \textwidth]{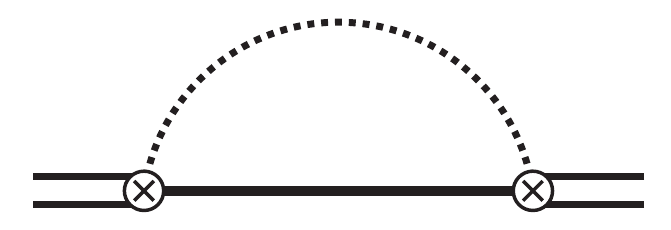}
\put(-320,20){a)}
\put(-135,20){b)}
\caption{\it One-loop self-energy diagrams for the singlet (a) and octet (b) field. The dotted line represents the scalar the double line the octet and the bold single line the singlet. The crosses symbolize the $\ord(r^0)$ singlet-octet-scalar interaction from Eq.~\ref{LUS}. The corresponding Feynman rules are found in Appendix~\ref{feynrules}.}
\label{1loopdiags}
\end{figure}

At one-loop level the relevant singlet and octet self-energy diagrams are depicted in Fig.~\ref{1loopdiags}. There are also two one-loop octet self-energy diagrams without an intermediate singlet propagator, which vanish however for $E = V_{o}$ and are not shown.\footnote{It is also possible to perform the calculation with small, but finite $E - V_{o}$ as a regulator for intermediate IR divergences, cf. Ref.~\cite{Pineda:2011db}. We have checked that the additional terms induced by this regulator cancel in our final results at one and two loops.} The singlet diagram in Fig.~\ref{1loopdiags}a has already been computed in Ref.~\cite{Pineda:2007kz}.
Together with the octet diagram in Fig.~\ref{1loopdiags}b we find
\begin{align}
 b_s^{(1)} &= 4\, \ln\! \Big( \frac{2 \Delta V}{\nu} \Big) - 4\,, 
\qquad
 b_o^{(1)} = 4\, \ln\! \Big( \frac{- 2 \Delta V}{\nu} \Big) - 4\,,\\
 c_s^{(1)} &= c_o^{(1)} = 4 \,.
\end{align}

The relevant two-loop diagrams and their individual results after integration in $d$ dimensions are shown in Appendix~\ref{twoloopdiags}. We have solved the integrals using integration by parts and partial fraction techniques. Expanding the respective self-energy expressions in $\epsilon = (4-d)/2$ and absorbing the UV divergences into the potential counterterms $\delta V_{s/o}$ yields
\begin{align}
 b_s^{(2)} =&\; 4\, C_A \ln^2\! \Big( \frac{2 \Delta V}{\nu} \Big) +  \Big(4 + \frac{4 \pi^2}{3}\Big) C_A \ln\! \Big( \frac{2 \Delta V}{\nu} \Big) 
 - \big(12-8 \zeta(3) \big) C_A - \frac{4 \pi^2}{3} C_F\,, \label{bs2}\\[0.2 cm]
b_o^{(2)} =&\; 4\, C_A \ln^2\! \Big( \frac{- 2 \Delta V}{\nu} \Big) +  \big( \Big(16-\frac{8 \pi^2}{3}\Big) C_A - 32\, C_F \big) \ln\! \Big( \frac{- 2 \Delta V}{\nu} \Big) +  \nn\\
 &+ \big(8 \zeta(3)- 36 + 4 \pi^2 \big)C_A  + \Big(64-\frac{4 \pi^2}{3} \Big) C_F \,, \label{bo2}\\[0.2 cm]
c_s^{(2)} =&\; \Big(4 + \frac{4 \pi^2}{3}\Big) C_A \,,\\[0.2 cm]
c_o^{(2)} =&\; \Big(16-\frac{8 \pi^2}{3}\Big) C_A - 32\, C_F \,.
\end{align}

In QCD, in contrast, nonzero ultrasoft contributions to the singlet and octet static energies and potential anomalous dimensions first occur at NNLO in the multipole expansion, i.e. at $\ord(r^2)$. Explicit results of state-of-the-art pNRQCD two-loop calculations at this order can be found in Ref.~\cite{Pineda:2011db}. They are potentially relevant for calculations of the ${\cal N}=4$ SYM static energy at $\ord(\alpha^5/r)$ or the ``ordinary'' Wilson loop without couplings of the static quarks to the scalar field.

\section{RG Evolution and Matching}
\label{sectionRG}

In order to perform the NLL resummation for the static potentials we have to solve the RGE's in Eqs.~\eqref{SingAnomDim} and \eqref{OctAnomDim}.
For finite $N_c$, i.e. nonvanishing octet potential, the Wilson coefficients $V_s$ and $V_o$ mix into each other. Their NLL RGE's are coupled due to the linear dependence on $\Delta V \equiv V_o - V_s$. Therefore we better rewrite them in matrix form:
\begin{align}
 \nu \frac{d}{d\nu} \bigg[\begin{array}{c} \! V_s \! \cr \! V_o \! \end{array}\bigg]
 &= \,M\, \bigg[\begin{array}{c} \! V_s \! \cr \! V_o \! \end{array}\bigg]\,,
\qquad
M := \bigg[\begin{array}{cc} \! -A \,&\, A \! \cr \! -B \,&\, B \! \end{array}\bigg]\,,
\label{MatrixRGE}
\end{align}
with the coefficients 
\begin{align}
 A &= -C_F \Big(\,4\, \frac{\alpha}{\pi} + \frac{4}{3} (3+\pi ^2 ) C_A\, \frac{\alpha^2}{\pi^2} \,\Big) \,,\\
 B &= \Big(\frac{C_A}{2} - C_F \Big) \Big(\, 4\, \frac{\alpha}{\pi} + \frac{8}{3} \left( (6-\pi ^2) C_A - 12 C_F\right)  \frac{\alpha^2}{\pi^2}   \,\Big)\,.
\end{align}
The solution to Eq.~\eqref{MatrixRGE} is given by
\begin{align}
\bigg[\begin{array}{c} \! V_s(\nu) \! \cr \! V_o(\nu) \! \end{array}\bigg]
 &= \,\exp\big[M \ln (r\nu)\big]\, \bigg[\begin{array}{c} \! V_s(1/r) \! \cr \! V_o(1/r) \! \end{array}\bigg]\,,
\label{MatrixSolution}
\end{align}
where $V_{s/o}(1/r) \equiv V_{s/o}(r; \nu\!=\!1/r)$ is understood. Carrying out the matrix exponentiation in Eq.~\eqref{MatrixSolution} we obtain
\begin{align}
\big[V_s(\nu)\big]^{\rm \sss NLL} = \frac1{B-A} \Big[ B\,V_s(1/r)-A\,V_o(1/r)  + A\, \Delta V(1/r)\, (r \nu)^{B-A}  \Big]\,, \label{VsNLL}\\
\big[V_o(\nu)\big]^{\rm \sss NLL} = \frac1{B-A} \Big[ B\,V_s(1/r)-A\,V_o(1/r)  + B\, \Delta V(1/r)\, (r \nu)^{B-A}  \Big]\,. \label{VoNLL}
\end{align}

What remains to be computed for the complete NLL solutions are the NLO matching coefficients $V_{s,o}(1/r)$.
The perturbative ${\cal N}=4$ SYM theory calculation of the tree-level and the one-loop (crossed) ladder diagrams yields the bare potentials~\cite{Pineda:2007kz}
\begin{align}
V_{s,B}&=-\frac{2C_Fg^2r^{2\epsilon}}{4\pi}\frac{1}{r}\frac{\Gamma[1/2-\epsilon]}{\pi^{1/2-\epsilon}}
\bigg\{1 -C_A g^2r^{2\epsilon} \frac{\Gamma[1/2-2\epsilon]\Gamma^2[-\epsilon]}{8\pi^{2-\epsilon}\Gamma[-2\epsilon]\Gamma[1/2-\epsilon]} \,+\, \ord(g^4) \bigg\}  \,, 
\label{Vs1loopBare}\\
V_{o,B}&=  \frac{(C_F - C_A/2)}{C_F}\,V_{s,B}  \,+\, \ord(g^6)\,.
\label{Vo1loopBare}
\end{align}
To extract the $\msb$ renormalized expressions we replace $g \to g \nu_s^\epsilon \exp[\epsilon\,(\gamma_E - \ln (4\pi))/2]$ in Eqs.~\eqref{Vs1loopBare} and~\eqref{Vo1loopBare}, subtract the respective potential one-loop counterterms and take the $\epsilon \to 0$ limit. In this procedure it is important to keep the $\ord(\epsilon)$ term of the LO $\msb$ expression for $\Delta V$, which is inserted in the counterterms $\delta V_{s/o}$. Otherwise we would miss a finite piece that must be subtracted from the bare potentials. Finally we set $\nu_s=1/r$ and find
\begin{align}
 V_s(1/r) &= -\frac{2 C_F\, \alpha}{r} \Big(1+ 2 C_A \gamma_E \frac{\alpha}{\pi} \,+\, \ord(\alpha^2) \Big)  \,, \label{VsmatchNLO}\\
 V_o(1/r) &= \frac{(C_A\!-\!2C_F)\, \alpha}{r} \Big(1+ 2 C_A \gamma_E \frac{\alpha}{\pi}  \,+\, \ord(\alpha^2)\Big) \label{VomatchNLO} \,.
\end{align}
Adding the ultrasoft one-loop terms of Eqs.~\eqref{EusSing} and~\eqref{EusOct} to Eqs.~\eqref{VsNLL} and~\eqref{VoNLL}, respectively, we can now write the static singlet and octet energies with NLL resummation as
\begin{align}
\big[E_s(r) \big]^{\rm \sss NLL} &=  \big[V_s(\nu)\big]^{\rm \sss NLL} -\frac{4 C_F \,\alpha}{\pi }  \big[\Delta V\big]^{\sss \rm LL} \Big( \ln \big( 2 \big[\Delta  V \big]^{\sss \rm LL} \nu^{-1} \big)-1 \Big)  \,, \label{EsNLL}\\
\big[E_o(r) \big]^{\rm \sss NLL} &=  \big[V_o(\nu)\big]^{\rm \sss NLL} + \frac{2 (C_A \!-\! 2 C_F) \alpha}{\pi } \big[\Delta V\big]^{\sss \rm LL} \Big( \ln \big(\!- 2 \big[\Delta  V \big]^{\sss \rm LL} \nu^{-1} \big)-1 \Big)\,.
\label{EoNLL}
\end{align}
These expressions have a residual $\nu$ dependence, since the ultrasoft logarithmic terms at $\ord(\alpha^3)$ or higher, e.g. the ones in Eqs.~\eqref{bs2} and~\eqref{bo2}, are not included. In order to render the higher order ultrasoft logarithms small we should set $\nu \sim \Delta V$ in Eqs.~\eqref{EsNLL} and~\eqref{EoNLL}.

Upon expansion in $\alpha$ Eqs.~\eqref{EsNLL} and~\eqref{EoNLL} yield the corresponding NLO expressions,
\begin{align}
\big[E_s(r) \big]^{\rm \sss NLO} &= -\frac{2 C_F\, \alpha}{r} \Big( 1 + \frac{2 C_A \alpha }{\pi} \big[ \ln(2C_A\alpha) + \gamma_E - 1 \big] \Big) \,, \\
\big[E_o(r) \big]^{\rm \sss NLO} &= \frac{(C_A\!-\!2C_F)\, \alpha}{r} \Big( 1 + \frac{2 C_A \alpha }{\pi} \big[ \ln(-2C_A\alpha) + \gamma_E - 1 \big] \Big) \label{EoNLO}\,,
\end{align}
where the $\nu$ dependence of the soft and ultrasoft contributions has canceled completely.
Moreover we can extract all LL ($\propto \alpha(\alpha\ln\alpha)^n/r$) and NLL ($\propto \alpha^2(\alpha\ln\alpha)^n/r$) terms from Eqs.~\eqref{EsNLL} and~\eqref{EoNLL} by choosing $\nu \sim \Delta V \sim C_A \alpha/r$. In particular we can determine the NNLO corrections to the static energies up to a real constant $c_{s/o}^{\rm \sss NNLO}$:
\begin{align}
\delta\big[E_s(r) \big]^{\rm \sss NNLO} &= -\frac{4 C_A^2 C_F \alpha^3}{\pi^2 r} \Big[ \ln^2(2 C_A\alpha)+ \Big(1+2 \gamma_E +\frac{\pi^2}{3} \Big) \ln (2 C_A\alpha) + c_s^{\rm \sss NNLO} \Big]\,,
\label{EsNNLOlogs}
   \\
\delta\big[E_o(r) \big]^{\rm \sss NNLO} &= \frac{2 C_A^2 (C_A\!-\!2C_F) \alpha^3}{\pi^2 r} \times  \label{EoNNLOlogs} \\
&\qquad \times \Big[ \ln^2(-2 C_A\alpha)+\Big(4 +2 \gamma_E -\frac{8 C_F}{C_A}-\frac{2 \pi ^2}{3}\Big) \ln (-2 C_A\alpha) + c_o^{\rm \sss NNLO} \Big]\,.\nn
\end{align}
We find the same results if we add also the two-loop ultrasoft contributions in Eqs.~\eqref{EusSing} and~\eqref{EusOct} to Eqs.~\eqref{EsNLL} and~\eqref{EoNLL}, respectively, replace $\big[\Delta V\big]^{\sss \rm LL}$ by
\begin{align}
 \big[\Delta V]^{\rm \sss NLO} &= \frac{C_A \alpha}{r} \Big(1 + \frac{2 C_A \alpha}{\pi} \big[\ln(r\nu)+\gamma_E \big] \Big) \,,
\end{align}
and expand in $\alpha$.
To fix the constants $c_{s/o}^{\rm \sss NNLO}$ only the presently unknown $\ord(\alpha^3)$ terms in the (scheme dependent) soft matching conditions Eqs.~\eqref{VsmatchNLO} and Eqs.~\eqref{VomatchNLO} are missing. The respective ultrasoft components are already included in Eqs.~\eqref{EusSing} and~\eqref{EusOct}.

Obviously the expressions for the octet static energy in Eqs.~\eqref{EoNLL}, \eqref{EoNLO} and~\eqref{EoNNLOlogs} have a nonzero imaginary part. It is related to the decay of the octet state into the singlet and massless quanta in the adjoint representation (gluons, scalars, Majorana fermions),
\begin{align}
 {\rm Im}[E_o(r)] = - \Gamma_o/2\,,
\end{align}
whereas the real part is the physical energy. Since for $\nu = \Delta V$ the potential coefficient $V_o$ (and $\Delta V$ itself) is real, we can determine the total decay width $\Gamma_o$ directly from the ultrasoft contribution in Eq.~\eqref{EusOct}. Note that every $\Delta V$ in Eq.~\eqref{EusOct} is accompanied by a positive imaginary infinitesimal due to causality: $\Delta V \to \Delta V + i\epsilon$, cf. the propagator~\eqref{SingProp}. We thus obtain
\begin{align}
\Gamma_o = \frac{4 \,\alpha^2}{r} \Big(\,1 \,+\, \frac{2 \alpha}{3 \pi}  \left[3 C_A \ln (2 \alpha  C_A)+\left(6+3 \gamma_E -\pi ^2\right) C_A-12 C_F\right]  \,+\, \ord(\alpha^2)\Big) \,,
\end{align}
where we have used that $C_A(C_A-2C_F)=1$. To put the energy and decay width of the octet quark-antiquark system into a physical context one actually has to allow for the (ultrasoft) interaction with some gluon and/or scalar background needed to form an overall color singlet state, cf. Ref.~\cite{Pineda:2011db} for the QCD case.
This is however outside the scope of this paper.

\section{The large-$N_c$ limit}
\label{sectionlargeNc}

This section addresses the fixed order calculation of the static singlet energy in the large-$N_c$ limit. Eq.~\eqref{EusSing} already contains the ultrasoft terms through $\ord(\lambda^3)$. For the complete NNLO static energy we however also need the $\ord(\lambda^3)$ term in the soft matching coefficient, Eq.~\eqref{VsmatchNLO}. This requires a soft two-loop calculation.
As a first step we have explicitly computed the two-loop ladder diagrams in perturbative ${\cal N}=4$ SYM theory.\footnote{In order to avoid pinch singularities we have only computed the maximally non-Abelian term $\propto C_A^2 C_F$ of the non-planar box (crossed ladder) diagrams. Although the total color factor of the non-planar diagrams vanishes in the large-$N_c$ limit only the maximally non-Abelian term survives in the sum of all two-loop ladder diagrams after subtraction of the Coulomb pinches due to exponentiation.} We have evaluated the relevant momentum-space two-loop integrals using the techniques described in Ref.~\cite{SchroederPhD} and afterwards performed the Fourier transformation to $d$-dimensional position-space. From the result we obtain the two-loop ladder contribution to be added to Eq.~\eqref{Vs1loopBare}. In total we find
\begin{align}
V_{s, lad}^{bare} =& -\frac{\lambda\,  r^{2 \epsilon}}{4 \pi}  \frac1{r} \frac{\Gamma (1/2-\epsilon)}{\pi^{1/2-\epsilon}} \, \bigg\{ 1
-\lambda\, r^{2\epsilon}\, \frac{\Gamma[1/2-2\epsilon]\Gamma^2[-\epsilon]}{8\pi^{2-\epsilon}\Gamma[-2\epsilon]\Gamma[1/2-\epsilon]} \nn \\[0.2 cm]
&-\lambda^2 r^{4 \epsilon}\, \frac{ 2 \pi  \Gamma(-3 \epsilon -1) \Gamma^3(1/2-\epsilon) + \Gamma(1/2 -3 \epsilon) \Gamma^3(-\epsilon ) }
{96 \pi ^{4-2\epsilon} (2 \epsilon +1) \Gamma(-3 \epsilon -1) \Gamma(1/2 -\epsilon)}
 \,+\, \ord(\lambda^3) \bigg\} \,.
\label{Vsladbare}
\end{align}
The octet static potential vanishes in the planar limit so that $\Delta V = -V_s$.


Since in our analysis we will treat ladder and non-ladder contributions separately, we also have to split the ultrasoft contributions and counterterms into two parts. This is easily possible, because full theory diagrams with a scalar or gluon ($A^0$) propagator connecting two points on the same static quark line cancel in the sum. Accordingly only the ultrasoft diagrams with direct interaction among the scalars, gluons and/or Majorana fermions, i.e. diagrams~\eqref{DiagNonLad1} and~\eqref{DiagNonLad2}, are of non-ladder type. Thus we have
\begin{align}
\delta E_{s,lad}^{us} =& 
-\Delta V \bigg\{ \frac{\lambda}{2 \pi ^2} \bigg[ \ln\! \Big(\frac{2 \Delta V}{\nu }\Big) -1 \bigg]
+
 \frac{\lambda^2}{8 \pi^4} \bigg[ \ln^2\!\Big(\frac{2 \Delta V}{\nu }\Big) + \ln\! \Big(\frac{2 \Delta V}{\nu}\Big) -3 \bigg]
+\ord(\lambda^3)  \bigg\} \nn\\
& + \ord(r^2) \,, \label{EusLad}\\
\delta E_{s,nonlad}^{us} =&
-\Delta V \bigg\{ \frac{\lambda^2}{8 \pi^4} \bigg[ \frac{\pi^2}{3}  \ln \!\Big(\frac{2 \Delta V}{\nu}\Big) -\frac{\pi ^2}{6} + 2 \zeta(3) \bigg] + \ord(\lambda^3)  \bigg\} + \ord(r^2) \label{EusNonLad}\,.
\end{align}
The associated counterterms read
\begin{align}
 \delta V_{s,lad} & = - \frac{\Delta V}{4 \pi^2} \Big[\, \frac{\lambda}{\epsilon} +  \frac{\lambda^2}{8 \pi^2 \epsilon^2}  +  \frac{\lambda^2}{8 \pi^2 \epsilon} +\ord(\lambda^3) \Big] 
+ \ord(r^2)\,, \label{deltaVslad}\\
 \delta V_{s,nonlad} &= - \frac{\Delta V }{96 \pi^2 } \Big[\, \frac{\lambda^2}{\epsilon} +\ord(\lambda^3) \Big] + \ord(r^2) \,.
\label{deltaVsnonlad}
\end{align}

To convert the bare expression Eq.~\eqref{Vsladbare} to the $\msb$ scheme we introduce the renormalization/factorization scale $\nu$ by replacing $\lambda \to \lambda \nu^{2\epsilon} \exp[\epsilon\,(\gamma_E - \ln(4\pi))]$ and subtract the counterterm $\delta V_{s,lad}$.
Note that now subleading terms of $\Delta V$ not only in $\epsilon$, but also in $\lambda$ must be taken into account in Eq.~\eqref{deltaVslad}. 
Before we can write down the NNLO result we therefore have to determine the NLO expression for $V_s=-\Delta V$, because the latter in turn is required in Eq.~\eqref{deltaVslad} as an input to reach NNLO precision. The outcome of this procedure is
\begin{align}
 V_{s,lad}(\nu) &= - \frac{\lambda}{4\pi\,r} \bigg\{ 1+ \frac{\lambda}{2 \pi^2} \ln(e^{\gamma_E}r\nu) + \frac{\lambda^2}{8 \pi^4} \bigg[ \ln^2(e^{\gamma_E}r\nu) + \ln(e^{\gamma_E}r\nu) - \frac12 -\frac{\pi^2}{12} \bigg] + \ord(\lambda^3) \bigg\} \,.
\label{VsladMsb}
\end{align}
Adding the ultrasoft ladder contribution of Eq.~\eqref{EusLad}, where again the NLO expression for $\Delta V$ related to the first two terms of Eq.~\eqref{VsladMsb}, must be inserted, we arrive at
\begin{align}
 E_{s,lad} =& - \frac{\lambda}{4\pi \, r}\bigg\{
1 + \frac{\lambda}{2 \pi^2} \bigg[\ln\! \Big( \frac{e^{\gamma_E}\lambda}{2\pi}\Big)-1 \bigg]  
+ \frac{\lambda^2}{8 \pi^4} \bigg[\ln^2 \!\Big( \frac{e^{\gamma_E}\lambda}{2\pi}\Big)+\ln\! \Big( \frac{e^{\gamma_E}\lambda}{2\pi}\Big) -\frac{\pi^2}{12}-\frac72 \bigg] \nn\\
 &+ \ord(\lambda^3) \bigg\}\,.
\label{Eslad}
\end{align}
This expression exactly agrees with the ladder result for the static energy in Ref.~\cite{Correa:2012nk}\,.

As for the non-ladder part of the static energy only the analog to Eq.~\eqref{VsladMsb} is missing. 
Its logarithmic term is however uniquely determined by requiring that the $\nu$-dependence exactly cancels the one of the ultrasoft non-ladder part in Eq.~\eqref{EusNonLad}. This is equivalent to the postulation that the IR divergence in the soft non-ladder calculation must match the ultrasoft ($1/\epsilon$) UV divergence in Eq.~\eqref{deltaVsnonlad}.
We can thus write
\begin{align}
 V_{s,nonlad}(\nu) &= - \frac{\lambda}{4\pi\,r} \bigg\{ \frac{\lambda^2}{8 \pi^4} \bigg[ \frac{\pi^2}{3} \ln(e^{\gamma_E}r\nu) + c^{\sss \rm NNLO,soft}_{nonlad} \bigg] + \ord(\lambda^3) \bigg\} \,.
\label{VsnonladMsb}
\end{align}
and together with Eq.~\eqref{EusNonLad},
\begin{align}
 E_{s,nonlad} =& - \frac{\lambda}{4\pi \, r}\bigg\{
\frac{\lambda^2}{8 \pi^4} \bigg[ \frac{\pi^2}{3} \ln\! \Big(\frac{e^{\gamma_E}\lambda}{2\pi}\Big) + c^{\sss \rm NNLO,soft}_{nonlad} - \frac{\pi ^2}{6} + 2 \zeta(3) \bigg]
+ \ord(\lambda^3) \bigg\}\,.
\label{Esnonlad}
\end{align}
We note that due to the additional factor $\pi^2/3$ the single non-ladder logarithm at $\ord(\lambda^3)$ is numerically dominant compared to the corresponding ladder logarithm.
The sum of the ladder and non-ladder logarithmic terms $\propto \lambda^3 (\ln \lambda)^n$ in Eqs.~\eqref{Eslad} and~\eqref{Esnonlad} consistently agrees with the large-$N_c$ limit of Eq.~\eqref{EsNNLOlogs}.
The real constant $c^{\sss \rm NNLO,soft}_{nonlad}$ can be obtained from the soft two-loop matching condition for the non-ladder potential $V_{s,nonlad}$ in the $\msb$ scheme, which is unknown at present.

A related soft two-loop calculation was performed in Ref.~\cite{Correa:2012nk}.
In contrast to our approach, however, the potential in Ref.~\cite{Correa:2012nk} was computed on the $S^3$ sphere, where the static quarks are separated by an angle $\delta$. The limit $\delta \to 0$ then yields the static potential in flat space. In this limit $\delta$ plays the role of the distance $r$ and at the same time acts as IR cutoff.
The (soft) results of Ref.~\cite{Correa:2012nk},
\begin{align}
V_{s,lad}^\delta &= - \frac{\lambda}{4\pi\,\delta} \bigg\{ 1 + \frac{\lambda}{2 \pi^2} \ln\! \Big(\frac{2 \delta}{e}\Big) + \frac{\lambda^2}{8 \pi^4} \bigg[\frac{\pi^3}{6\delta} + \ln^2\!\Big(\frac{2 \delta}{e}\Big) 
+ \ln\! \Big(\frac{2 \delta}{e}\Big) - \frac32 -\frac{\pi^2}{4} \bigg] + \ord(\lambda^3) \bigg\}\,,   
\label{Vsladdelta}\\
V_{s,nonlad}^\delta &= - \frac{\lambda}{4\pi\,\delta} \frac{\lambda^2}{8 \pi^4} \bigg[ \frac{\pi^2}{3} \ln\! \Big(\frac{2 \delta}{e}\Big) + \frac{\pi^2}{3} + \frac{9 \zeta(3)}{4} \bigg] + \ord(\lambda^3) \bigg\}\,,
\label{Vsnonladdelta}
\end{align}
are therefore expressed in a (cutoff) scheme different from our $\msb$ scheme.
Note that Eq.~\eqref{Vsladdelta} contains a power-like IR divergence of the form $1/\delta^2$. It is canceled exactly by the corresponding ultrasoft expression in the $\delta$-scheme~\cite{Correa:2012nk} and does not have an equivalent in Eq.~\eqref{VsladMsb}, because the latter has been derived using dimensional regularization.

It is tempting to try to translate Eq.~\eqref{Vsnonladdelta} to the $\msb$ scheme in order to fix the constant $c^{\sss \rm NNLO,soft}_{nonlad}$, or alternatively translate Eq.~\eqref{EusNonLad} to the $\delta$-scheme, by devising a dictionary for the logarithmic terms (in addition to the simple replacement rule $1/\delta \leftrightarrow 1/r$) based on the knowledge of the respective ladder results in both schemes.
We however refrain from doing so since the conversion of multi-loop results between different regularization schemes, in particular cutoff schemes and dimensional regularization, is a delicate issue and naive replacement rules can be misleading. Given the involved loop structure in our case a reliable translation prescription would require a dedicated study which goes beyond the scope of this paper. On the other hand a direct ($\msb$) computation of $c^{\sss \rm NNLO,soft}_{nonlad}$ from soft non-ladder two-loop quark-antiquark interaction diagrams in flat space is extensive, but certainly feasible with present Feynman diagram technology. This work is left for the future.

Summarizing the fixed order results for the static energy in the large-$N_c$ limit we have
\begin{align}
E_s(r) =& - \frac{\lambda}{4\pi \, r}\Bigg\{
1 + \frac{\lambda}{2 \pi^2} \bigg[\ln\! \Big( \frac{e^{\gamma_E}\lambda}{2\pi}\Big)-1 \bigg]  \nn\\[1 ex]
& \hspace{10.3 ex}+ \frac{\lambda^2}{8 \pi^4} \bigg[\ln^2 \!\Big( \frac{e^{\gamma_E}\lambda}{2\pi}\Big)+ \Big(1+\frac{\pi^2}{3} \Big) \ln\! \Big( \frac{e^{\gamma_E}\lambda}{2\pi}\Big) -\frac{\pi^2}{4}-\frac72 + 2 \zeta(3) + c^{\sss \rm NNLO,soft}_{nonlad} \bigg] \nn\\[1 ex]
& \hspace{10.3 ex}+ \frac{\lambda^3}{48 \pi^6} \bigg[\ln^3 \!\Big( \frac{e^{\gamma_E}\lambda}{2\pi}\Big)+ (6+\pi^2) \ln^2\! \Big( \frac{e^{\gamma_E}\lambda}{2\pi}\Big)+ c^{\sss \rm NNLL}_1 \ln\! \Big( \frac{e^{\gamma_E}\lambda}{2\pi}\Big) + c^{\sss \rm N^3LO} \bigg] \nn\\[1 ex]
& \hspace{10.3 ex}+ \frac{\lambda^4}{384 \pi^8} \bigg[\ln^4 \!\Big( \frac{e^{\gamma_E}\lambda}{2\pi}\Big)+ 2(7+\pi^2) \ln^3\! \Big( \frac{e^{\gamma_E}\lambda}{2\pi}\Big)+ c^{\sss \rm NNLL}_2 \ln^2\! \Big( \frac{e^{\gamma_E}\lambda}{2\pi}\Big) \nn\\[1 ex]
& \hspace{20 ex}+ c^{\sss \rm N^3LL}_1 \ln\! \Big( \frac{e^{\gamma_E}\lambda}{2\pi}\Big) + c^{\sss \rm N^4LO} \bigg]
\;+\; \ord(\lambda^5) \Bigg\}\,.
\label{EslargeNc}
\end{align}
Here we have parametrized the currently unknown terms by coefficients $c_i^{\sss X}$ and explicitly expanded out the LL and NLL large-$N_c$ logarithms at $\ord(\lambda^4)$ and $\ord(\lambda^5)$ from Eq.~\eqref{EsNLL} by choosing $\nu \sim \lambda/(4\pi r)$ for illustration. All the $\pi^2$-enhanced NLL logarithms in Eq.~\eqref{EslargeNc} come from non-ladder diagrams, all other LL and NLL logarithms are generated by ladder diagrams.

The one-loop logarithm in the first line of Eq.\eqref{EslargeNc} was first determined by Erickson et al. in Ref.~\cite{Erickson:1999qv}, the one-loop constant and all LL logarithms by Pineda in Ref.~\cite{Pineda:2007kz}. The $\ord(\lambda^3)$ ladder result, Eq.~\eqref{Eslad}, was first calculated by Correa et al. in Ref.\cite{Correa:2012nk}. Bykov and Zarembo computed in Ref.~\cite{Bykov:2012sc} the leading logarithmic correction to the ladder result, which agrees with our $\ord(\lambda^3)$ non-ladder logarithm in Eq.~\eqref{Esnonlad}. They however considered their result as incomplete and assumed that there are also other terms $\propto \lambda^3 \ln \lambda$ at higher orders in their expansion about the ladder result, which they had not computed.
We have shown that this is not the case and Eq.~\eqref{EslargeNc} contains the complete LL and NLL, ladder as well as non-ladder, contributions. The remaining terms displayed in Eq.~\eqref{EslargeNc} are also new.

As already mentioned the NNLO constant, $c^{\sss \rm NNLO,soft}_{nonlad}$, can be computed in terms of soft, i.e. full theory, two-loop non-ladder diagrams. The NNLL coefficients $c^{\sss \rm NNLL}_i$ require the UV-divergences of the ultrasoft three-loop self-energy at $\ord(r^0)$ and the ultrasoft one-loop self-energy at $\ord(r^2)$~\cite{Pineda:2007kz}, whereas for the determination of $c^{\sss \rm N^3LO}$ the respective ultrasoft finite parts in addition to the soft three-loop calculation are needed. Finally for the complete N$^4$LO static energy four-loop soft and ultrasoft calculations have to be performed, where the ultrasoft four-loop $\ord(r^0)$ and two-loop $\ord(r^2)$ UV-divergences suffice to fix the N$^3$LL coefficients $c^{\sss \rm N^3LL}_i$.

\section{Conclusion}
\label{sectionConclusion}

The ${\cal N}=4$ SYM potential (energy) between two static sources in the fundamental representation of $SU(N_c)$ is composed of soft and ultrasoft contributions. Using the effective theory proposed in Ref.~\cite{Pineda:2007kz} we have calculated the respective ultrasoft contributions for the singlet and adjoint color state at two loops.

The main results of the paper are the NLL expressions for the ${\cal N}=4$ SYM singlet and adjoint static energy in Eqs.~\eqref{EsNLL} and~\eqref{EoNLL} as well as the large-$N_c$ fixed order expression for the singlet static energy in Eq.~\eqref{EslargeNc}. For the renormalization scale choice $\nu \sim C_A \alpha/r$ the former two expressions contain the complete resummation of logarithmic terms $\propto \alpha(\alpha \ln \alpha)^n/r$ and $\propto \alpha^2 (\alpha \ln \alpha)^n/r$ for arbitrary $N_c$. The latter expression includes the contributions from all relevant diagrams to $\ord(\lambda^3)$ except for an unknown constant. This constant can be determined from a purely perturbative (soft matching) computation of the two-loop non-ladder static quark-antiquark interaction diagrams in flat space. We leave this calculation for the future.

\acknowledgments{
I am very grateful to Antonio Pineda for carefully reading the manuscript and many helpful comments.
}
 
\newpage

\appendix

\section{EFT Feynman Rules}
\label{feynrules}

This appendix lists the momentum space Feynman rules associated with the $\ord(r^0)$ EFT Lagrangian in Eq.~\eqref{LUS}.

\subsection{Singlet/octet vertices:}
\vspace{- 3 ex}

\begin{align}
\raisebox{-2 ex}{\includegraphics[width=0.22 \textwidth]{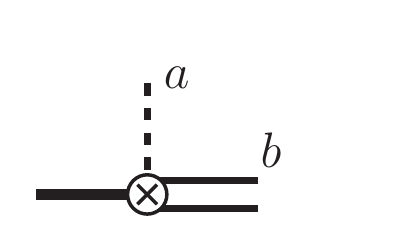}}&\hspace{-5 ex}:\quad -2\,i\,g\, V_{\phi}\, \sqrt{\frac{T_F}{N_c}}\, \delta^{ab}    \\
\raisebox{-2 ex}{\includegraphics[width=0.22 \textwidth]{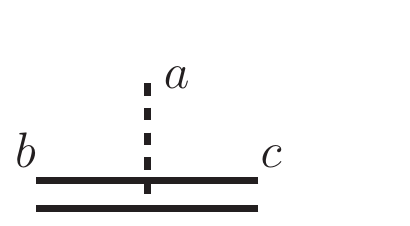}}&\hspace{-5 ex}:\quad - i\, g\, V_{\Phi_O}\, d^{abc}     \\
\raisebox{-2 ex}{\includegraphics[width=0.22 \textwidth]{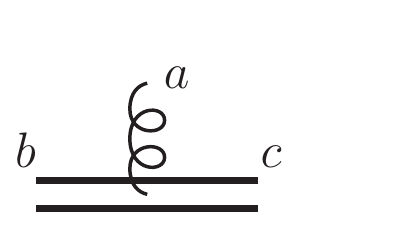}}&\hspace{-5 ex}:\quad g\,f^{abc}
\end{align}
\\[-5 ex]

\subsection{Singlet/octet propagators:}
\vspace{- 4 ex}

\begin{align}
\raisebox{-0.5 ex}{\includegraphics[width=0.15 \textwidth]{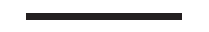}}& :\quad  \frac{i}{k^0-V_s + i\epsilon} \label{SingProp}\\[2 ex]
\raisebox{-1.5 ex}{\includegraphics[width=0.15 \textwidth]{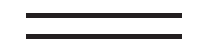}}& :\quad  \frac{i}{k^0-V_o + i\epsilon}
\end{align}
\\[-2 ex]

Conventionally a left-right flow of the ultrasoft four-momentum $k^\mu$ is  understood.
The dotted line represents the scalar the double line the octet, the bold single line the singlet and the spiral line the zero component of the gluon field ($A^0$). The letters $a,b,c$ denote color indices in the adjoint representation. The $f^{abc}$ and $d^{abc}$ are the totally antisymmetric and symmetric structure constants of $SU(N_c)$, respectively.

\newpage

\section{EFT Diagrams}
\label{twoloopdiags}

In the following amplitudes we set $V_{\phi}=V_{\Phi_O}=1$ and $T_F=1/2$.
The expression for the scalar one-loop self-energy subgraph (regularized by dimensional reduction) in the diagrams~\eqref{DiagNonLad2} and~\eqref{OctScalarSelf} can be found in Ref.~\cite{Erickson:2000af}.

\subsection{Two-loop singlet self-energy diagrams:}
\begin{align}
 X_s&=-\frac{i}{16 \pi^d} \, g^4 \,C_F\, \Delta V^{2 d-7}
\end{align}

\begin{align}
X_s^{-1}\raisebox{-0.5 ex}{\includegraphics[width=0.25 \textwidth]{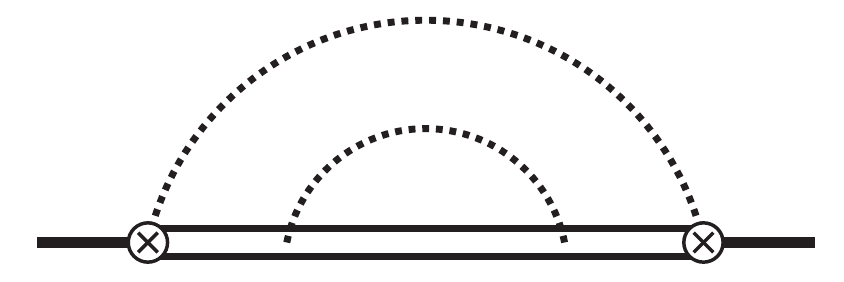}}&=
-\frac{\sqrt{\pi } C_d 2^{5-d} \Gamma (7-2 d) \Gamma \left(\frac{d}{2}-1\right) \Gamma (d-4)}{\Gamma \left(\frac{d-1}{2}\right)}
 \\[1 ex]
X_s^{-1}\raisebox{-0.5 ex}{\includegraphics[width=0.25 \textwidth]{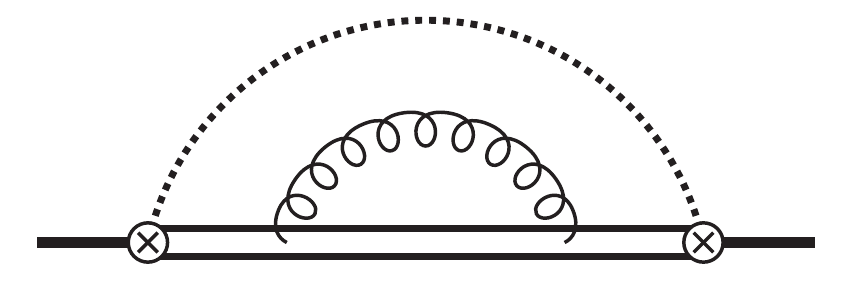}}&=
\frac{\sqrt{\pi } C_A 2^{5-d} \Gamma (7-2 d) \Gamma \left(\frac{d}{2}-1\right) \Gamma (d-4)}{\Gamma \left(\frac{d-1}{2}\right)}
 \\[1 ex]
X_s^{-1}\raisebox{-0.5 ex}{\includegraphics[width=0.25 \textwidth]{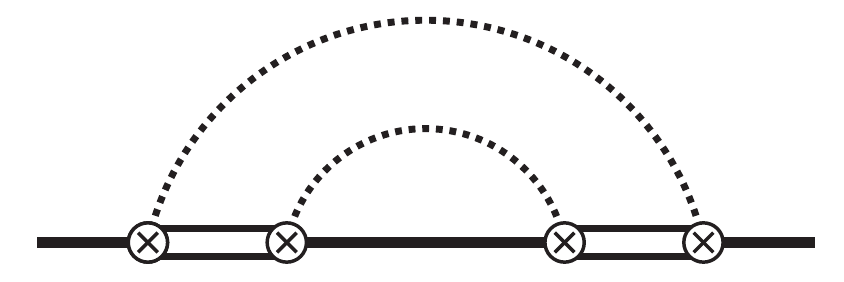}}&= 
\frac{\sqrt{\pi } 2^{7-d} (C_A-2 C_F) \Gamma (1-2 d) \Gamma (4-d) \Gamma \left(\frac{d}{2}-1\right) \Gamma (2 d)}{\Gamma \left(\frac{d-1}{2}\right)}
\\[4 ex]
X_s^{-1}\raisebox{-3 ex}{\includegraphics[width=0.25 \textwidth]{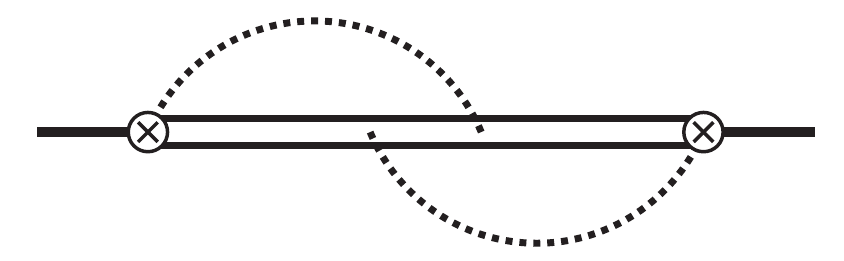}}&= 
\frac{8 C_d \Gamma \left(\frac{d}{2}-1\right)\! \left[\Gamma (6-2 d) \Gamma \left(\frac{d}{2}-1\right)\!+\Gamma (2-d) \Gamma(4-d) \Gamma \left(\frac{d}{2}\right)\right]}{d-3}
\\[-1 ex]
X_s^{-1}\raisebox{-3 ex}{\includegraphics[width=0.25 \textwidth]{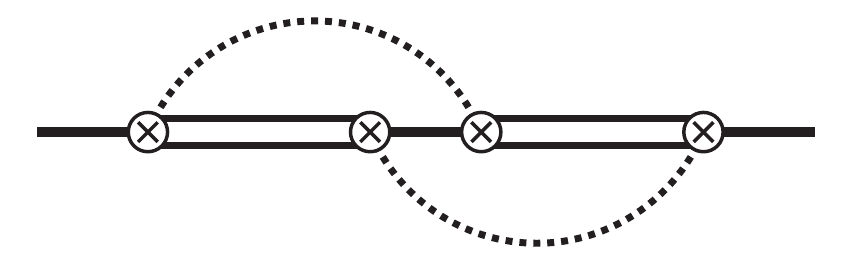}}&= 
\frac{\sqrt{\pi } 2^{5-d} (C_A\!-\!2 C_F)\, \Gamma (3-d) \Gamma \left(\frac{d}{2}-1\right)}{\Gamma \left(\frac{d-1}{2}\right)} \times \nn\\
& \qquad \times \big[\Gamma (1-d) \Gamma (d)+2 \Gamma (1-2d) \Gamma (2 d) \big]
 \\[1 ex]
X_s^{-1}\raisebox{-0.5 ex}{\includegraphics[width=0.25 \textwidth]{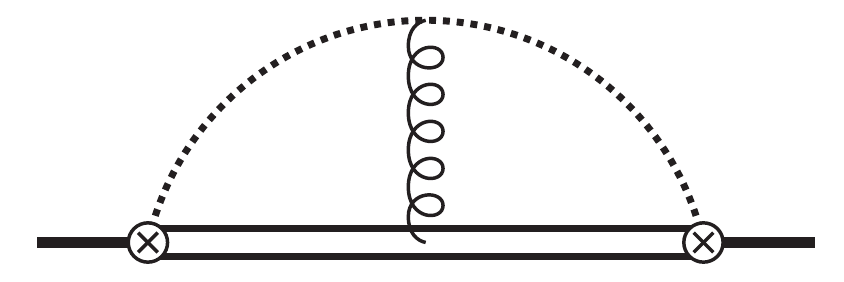}}&=
C_A [(6 d-20) \Gamma (6-2 d)+\Gamma (3-d) \Gamma (5-d)] \Gamma^2 \Big(\frac{d}{2}-2\Big)
    \label{DiagNonLad1}\\[1 ex]
X_s^{-1}\raisebox{-0.5 ex}{\includegraphics[width=0.25 \textwidth]{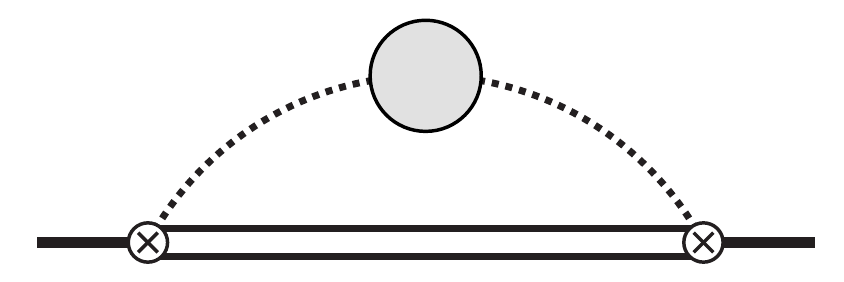}}&=
2 C_A (d-4) \Gamma (6-2 d) \Gamma^2 \Big(\frac{d}{2}-2\Big)
    \label{DiagNonLad2}
\end{align}

\subsection{Two-loop octet self-energy diagrams:}
(Left-right mirror graphs are included in the expressions below.)

\begin{align}
 X_o&= -\frac{i}{16 \pi^d}\, g^4 \Big(\frac{C_A}{2} - C_F\Big) (-\Delta V)^{2 d-7}
\end{align}

\begin{align}
X_o^{-1}\raisebox{-0.5 ex}{\includegraphics[width=0.25 \textwidth]{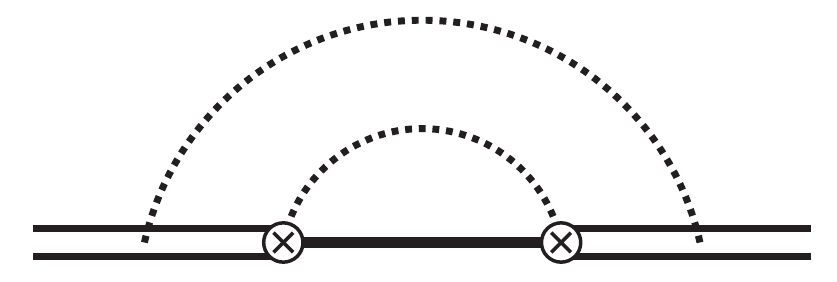}}&= X_s^{-1}\raisebox{-0.5 ex}{\includegraphics[width=0.25 \textwidth]{Figs/N4SYM2loopDiagB2.pdf}}
\\[2 ex]
X_o^{-1}\raisebox{-0.5 ex}{\includegraphics[width=0.25 \textwidth]{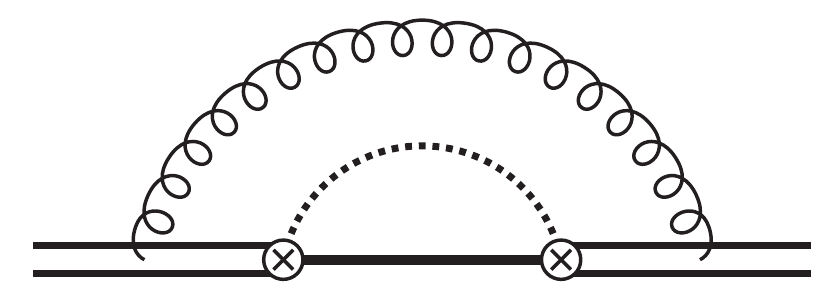}}&= X_s^{-1}\raisebox{-0.5 ex}{\includegraphics[width=0.25 \textwidth]{Figs/N4SYM2loopDiagB3.pdf}}
\\[2 ex]
X_o^{-1}\raisebox{-0.5 ex}{\includegraphics[width=0.25 \textwidth]{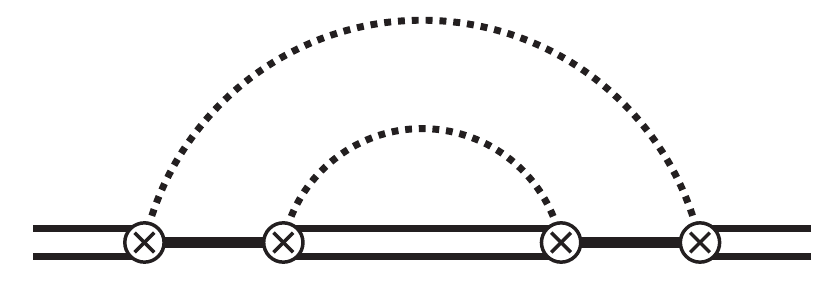}}&= \frac{2 C_F}{C_A-2 C_F}\; X_s^{-1}\raisebox{-0.5 ex}{\includegraphics[width=0.25 \textwidth]{Figs/N4SYM2loopDiagB4.pdf}}
\\[4 ex]
X_o^{-1}\raisebox{-3 ex}{\includegraphics[width=0.25 \textwidth]{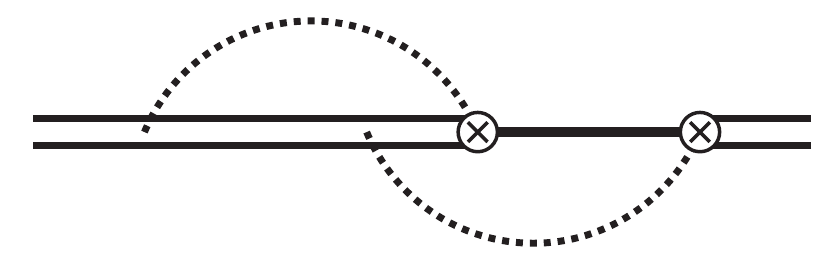}}&= \frac{\sqrt{\pi } C_d 2^{6-d} \Gamma (1-2 d) \Gamma (3-d) \Gamma \left(\frac{d}{2}-1\right) \Gamma (2 d)}{\Gamma
   \left(\frac{d-1}{2}\right)}
\\[3 ex]
X_o^{-1}\raisebox{-3 ex}{\includegraphics[width=0.25 \textwidth]{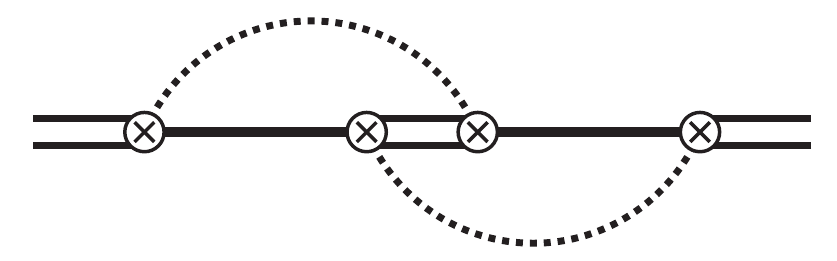}}&= X_s^{-1}\raisebox{-3 ex}{\includegraphics[width=0.25 \textwidth]{Figs/N4SYM2loopDiagB7.pdf}}
\\[2 ex]
X_o^{-1}\raisebox{-0.5 ex}{\includegraphics[width=0.25 \textwidth]{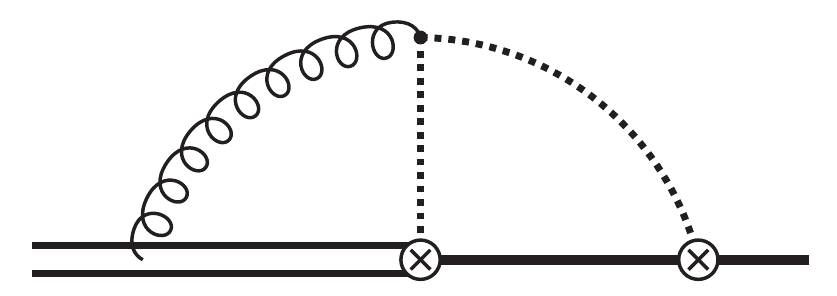}}&= 
\frac{\sqrt{\pi } C_A 2^{5-d} \Gamma \left(1-\frac{d}{2}\right) \Gamma \left(\frac{d}{2}\right) }
{(d-4) \Gamma \left(3-\frac{d}{2}\right) \Gamma (d-2) \Gamma \left(\frac{d-1}{2}\right)} \times  \\
&\hspace{-18 ex}\times \big[(3 d-10) \Gamma (6-2 d) \Gamma (d-2)^2  +   (d-4) (d-3) \Gamma (1-2 d) \Gamma (1-d) \Gamma (d) \Gamma (2 d)\big] \nn
\\[2 ex]
X_o^{-1}\raisebox{-0.5 ex}{\includegraphics[width=0.25 \textwidth]{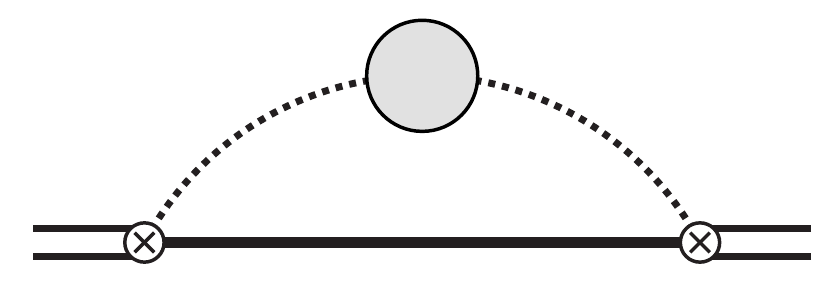}}&= X_s^{-1}\raisebox{-0.5 ex}{\includegraphics[width=0.25 \textwidth]{Figs/N4SYM2loopDiagB5.pdf}}
\label{OctScalarSelf}
\end{align}

\newpage
%

\bibliographystyle{JHEP}
\addcontentsline{toc}{chapter}{Bibliography}
\bibliography{MeiBib}

\end{document}